\documentclass[aps,pre,superscriptaddress,reprint]{revtex4-2}
\usepackage{amsmath}
\usepackage{amsfonts}
\usepackage{microtype}
\usepackage{graphicx}
\usepackage{xcolor}

\newcommand{\dpar}[1]{\left(#1\right)}
\newcommand{\dsqr}[1]{\left[#1\right]}
\newcommand{\dcur}[1]{\left\{#1\right\}}

\newcommand{\R}{\mathbb{R}}
\newcommand{\mc}{\mathcal}

\begin{document}

\title{Stationary solution and $H$ theorem for a generalized Fokker-Planck equation}

\author{Max Jauregui}
\email[E-mail: ]{8jauregui@gmail.com}
\affiliation{Departamento de Matem\'atica, Universidade Estadual de Maring\'a, Av. Colombo, 5790 CEP 87020-900 - Maring\'a - PR - Brazil}

\author{Anna L. F. Lucchi}
\affiliation{Departamento de F\'isica, Universidade Estadual de Maring\'a, Av. Colombo, 5790 CEP 87020-900 - Maring\'a - PR - Brazil}
\affiliation{National Institute of Science and Technology for Complex Systems, Rua Xavier Sigaud 150, 22290-180 Rio de Janeiro, Brazil}

\author{Jean H. Y. Passos}
\affiliation{Departamento de F\'isica, Universidade Estadual de Maring\'a, Av. Colombo, 5790 CEP 87020-900 - Maring\'a - PR - Brazil}
\affiliation{National Institute of Science and Technology for Complex Systems, Rua Xavier Sigaud 150, 22290-180 Rio de Janeiro, Brazil}

\author{Renio S. Mendes}
\affiliation{Departamento de F\'isica, Universidade Estadual de Maring\'a, Av. Colombo, 5790 CEP 87020-900 - Maring\'a - PR - Brazil}
\affiliation{National Institute of Science and Technology for Complex Systems, Rua Xavier Sigaud 150, 22290-180 Rio de Janeiro, Brazil}

\begin{abstract}
  We investigate a family of generalized Fokker-Planck equations that contains Richardson and porous media equations as members. Considering a confining drift term that is related to an effective potential, we show that each equation of this family has a stationary solution that depends on this potential. This stationary solution encompasses several well-known probability distributions. Moreover, we verify an $H$ theorem for the generalized Fokker-Planck equations using free-energy-like functionals. We show that the energy-like part of each functional is based on the effective potential and the entropy-like part is a generalized Tsallis entropic form, which has an unusual dependence on the position and can be related to a generalization of the Kullback-Leibler divergence.
    We also verify that the optimization of this entropic-like form subjected to convenient constraints recovers the stationary solution. The analysis presented here includes several studies about $H$ theorems for other generalized Fokker-Planck equations as particular cases.

\end{abstract}

\maketitle

\section{Introduction}

The Gaussian and exponential distributions frequently appear in the study of many physical systems. In fact, the Gaussian distribution is in a privileged position, since it is the limit distribution for the sum of independent random variables with finite variance according to the central limit theorem~\cite{Durrett}. On the other hand the exponential distribution is also of great importance in statistical mechanics, since it is at the core of the Boltzmann-Gibbs weight~\cite{GallavottiSM}.

In addition to the Gaussian and the exponential distributions, others have been employed in the analysis of many natural, social and artificial systems. For instance, the Laplace distribution appears in the analysis of virtual and traditional currency exchange rates~\cite{10.1371/journal.pone.0220070}; a generalization of this distribution, sometimes called exponential power or Subbotin distribution (density proportional to $\exp(-\beta|x|^r)$), occurs as a generalized law of errors~\cite{Subbotin1923}. In medicine, a study about the incidence of the most prevalent cancer types in relation to the age of the patients employs the gamma distribution~\cite{Belikov2017}, which contains the exponential one as a special case. Another generalization of the exponential distribution, called Weibull distribution, has several applications in physics and economy~\cite{Laherrere1998}.

All the distributions mentioned in the last paragraph involve the exponential function in their definitions. In addition to them, power law distributions commonly appear in the study of complex systems. In this direction, a few but representative examples are listed in the following. The $q$-exponential distribution (density proportional to $[1-(1-q)\beta x]^{1/(1-q)}$), which is a generalization of the exponential distribution related to the nonextensive statistical mechanics~\cite{Tsallis}, is used to describe the distribution of scientific citations~\cite{Anastasiadis2010}. The distribution of price returns of the Standard \& Poor's 500 stock market~\cite{PhysRevE.99.062313} is well fitted by a $q$-Gaussian distribution (density proportional to $[1-(1-q)\beta x^2]^{1/(1-q)}$), which is a generalization of the Gaussian one. The $q$-gamma distribution (density proportional to $x^{\alpha-1}[1-(1-q)\beta x]^{1/(1-q)}$) is a generalization of the gamma distribution that is applied to characterize the volume distribution in diverse stock markets in the high-frequency scale~\cite{OsorioBorlandTsallis2004,Souza_etal2006,Cortines_2008,DUARTEQUEIROS201624}. Finally, the number of active cases in the first months of the COVID-19 pandemic has been modeled by a generalization of the $q$-gamma distribution (density proportional to $x^a[1-(1-q)\beta x^r]^{1/(1-q)}$)~\cite{10.3389/fphy.2020.00217}.


In the study of diffusion processes, a well known fact is that the fundamental solution of the usual linear diffusion equation is a Gaussian distribution. In addition, several works have considered generalized diffusion equations which have solutions related to non-Gaussian distributions; for instance, Subbotin and $q$-Gaussian ones. These generalizations usually involve the use of a variable diffusion coefficient~\cite{doi:10.1098/rspa.1926.0043,PhysRevE.94.062117}, the addition of nonlinear terms~\cite{PlastinoPlastino1995,PhysRevE.63.030101,PhysRevE.94.062117,PhysRevE.99.042141} or the employment of fractional derivatives~\cite{Metzler_2004,doi:10.1142/S0218127408021877,SILVA200765}.
In the context of probability distributions, the linear diffusion equation can be interpreted as a Fokker-Planck equation~\cite{Risken,PlastinoPlastino1995}. If the drift term is confining and proportional to $x$, this equation has a stationary solution which is a Gaussian distribution~\cite{Risken}. If we consider more general diffusion equations and, consequently, generalized Fokker-Planck equations, we could expect to obtain all the well-known distributions mentioned before as stationary solutions. Furthermore, the relaxation towards stationary solutions has been investigated in several studies in connection with $H$ theorems using free-energy-like functionals~\cite{Shiino2001,KANIADAKIS2001283,Frank,SchwammleCuradoNobre2007,SchwammleNobreCurado2007,PhysRevE.94.062117,Casas_2019}.
These functionals are constituted by an energy-like part and an entropy-like one. An important result of this kind of study is the possible emergence of new types of internal energies and of entropic forms. In particular, these new entropic forms and related distributions may find applications, for instance, in the study of complex systems.

Despite the recent advances in the study of nonlinear Fokker-Planck equations and $H$ theorems associated with them~\cite{SchwammleCuradoNobre2007,SchwammleNobreCurado2007,PhysRevE.94.062117,Manita_etal2015,Barbu2016,CoghiGess2019}, additional attention should be paid to obtaining solutions, new entropic forms and a connection between them. In this direction,
we will investigate here a family of generalized Fokker-Planck equations (see Section~\ref{sec.Eq}) whose stationary solution~(see Section~\ref{sec:stat}) contains all the distributions mentioned before as particular cases. We will show that an effective potential, depending on the diffusion coefficient, could appear in the expression of the stationary solution.
Moreover, we will verify an $H$ theorem (see Section~\ref{sec:Hthm}) considering a free-energy-like functional composed by an internal energy, which is a mean of the effective potential, and a generalized Tsallis entropic form. We will reveal that this entropic form, which depends on the position, can be connected to a generalization of the Kullback-Leibler divergence. In connection with a maximum entropy principle, we will consider the possibility of decomposing the free-energy-like functional into other internal energies and entropic forms. We will show that the optimization of the generalized Tsallis entropic form subjected to convenient constraints recovers the stationary solution, which does not happen, in general, for other entropic forms. Finally, concluding remarks are given in Section~\ref{sec:conc}. 


\section{A family of generalized Fokker-Planck equations}
\label{sec.Eq}

Motivated by the similitude between Fokker-Planck and diffusion equations, we will start our study of a generalized Fokker-Planck equation reviewing some aspects of diffusion equations related to well-known distributions.
%
In this direction, we firstly consider the time-dependent generalized $q$-gamma distribution
\begin{equation}
  \label{solution}
  \rho(x,t)=N\frac{|x|^a}{t^{(a+1)s/r}}\dpar{1-A\frac{|x|^r}{t^s}}_+^c\,,
\end{equation}
where $(x)_+=\max\{x,0\}$, $a$ and $s$ are arbitrary real constants, $c\ne 0$, $r\ne 0$, $A\ne 0$ and $N>0$ is a normalization constant (obtained from $\int_{-\infty}^{+\infty} \rho(x,t)\,dx=1$). We note that this distribution encompasses the possibility of power law behavior for $x\to 0$ and for $x\to\pm\infty$.
In addition, Eq.~(\ref{solution}) contains a function that involves the exponential function as a particular case. Indeed, choosing $A=\beta/c$, $\beta>0$, and taking the limit $c\to+\infty$, we obtain
\begin{equation}
  \label{solution.par}
  \bar\rho(x,t)=\lim_{\substack{c\to+\infty\\A=\beta/c}}\rho(x,t)=\frac{|r|\beta^{\ell}}{2\Gamma(\ell)}\frac{|x|^a}{t^{(a+1)s/r}}\exp\dpar{-\beta\frac{|x|^r}{t^s}}\,,
\end{equation}
 where $\ell=(a+1)/r$ and $\Gamma(\cdot)$ stands for the gamma function. Exceptionally, we note that, if $r<0$, the function $\bar\rho(x,t)$ does not have a power-law asymptotic behavior near the origin.

The functions $\rho(x,t)$ and $\bar\rho(x,t)$, defined by Eqs.~(\ref{solution}) and~(\ref{solution.par}), yield time-dependent solutions of several diffusion equations after choosing appropriate values for their parameters. For instance, considering $a=0$, $r=2$, $s=1$, $\beta=1/4D$, $D>0$ in Eq.~(\ref{solution.par}), we have
\begin{equation}
  \label{usual.sol}
  \bar\rho(x,t)=\frac{1}{\sqrt{4\pi Dt}}\exp\dpar{-\frac{x^2}{4Dt}}\,,
\end{equation}
which is the fundamental solution of the usual linear diffusion equation
\begin{equation}
  \label{usual}
  \frac{\partial\bar\rho}{\partial t}=D\frac{\partial^2 \bar\rho}{\partial x^2}\,.
\end{equation}
A generalization of Eq.~(\ref{usual.sol}) can also be obtained from Eq.~(\ref{solution.par}) by not fixing the value of the parameter $r$. In this case we obtain
\begin{equation}
  \label{Richardson.sol}
  \bar\rho(x,t)=\frac{|r|\beta^{1/r}}{2\Gamma(1/r)t^{1/r}}\exp\dpar{-\beta\frac{|x|^r}{t}}\,,
\end{equation}
which is a solution of the diffusion equation
\begin{equation}
  \label{Richardson}
  \frac{\partial \bar\rho}{\partial t}=D\frac{\partial}{\partial x}\dsqr{|x|^{2-r}\frac{\partial \bar\rho}{\partial x}}\,,
\end{equation}
where $D=1/r^2\beta$. Equation~(\ref{Richardson}) was originally proposed by Richardson~\cite{doi:10.1098/rspa.1926.0043}, who considered $r=2/3$.
We note that Eqs.~(\ref{Richardson}) and~(\ref{Richardson.sol}) recover respectively Eqs.~(\ref{usual}) and~(\ref{usual.sol}), if $r=2$.

Considering $a=0$, $c=1/(1-q)$, $r=2$, $s=2/(3-q)$ and $A=(1-q)\beta$ in Eq.~(\ref{solution}), where $q<2$ and $\beta>0$, we obtain
\begin{equation}
  \label{porous.sol}
  \rho(x,t)=\frac{N}{t^{1/(3-q)}}\dsqr{1-(1-q)\beta\frac{x^2}{t^{2/(3-q)}}}_+^{1/(1-q)}\,.
\end{equation}
We can verify that this function, which is a $q$-Gaussian distribution~\cite{Tsallis1988}, is a solution of the following nonlinear diffusion equation~\cite{PlastinoPlastino1995}
\begin{equation}
  \label{porous}
  \frac{\partial \rho}{\partial t}=D\frac{\partial^2 \rho^{2-q}}{\partial x^2}\,,
\end{equation}
where $D=N^{q-1}/[2(2-q)(3-q)\beta]$. Equation~(\ref{porous}) is usually called the porous media equation~\cite{Muskat,refId0,Aronson1986}, which appears, for instance, in the discussion of the percolation of gases through porous media~\cite{Muskat}, thin saturated regions in porous media~\cite{Polunarinova} and thin liquid films spreading under gravity~\cite{buckmaster_1977}. Straightforwardly, we can verify that Eqs.~(\ref{porous.sol}) and~(\ref{porous}) recover Eqs.~(\ref{usual.sol}) and~(\ref{usual}) if $q\to 1$.


Up to now, all the functions that we have shown as particular cases of Eqs.~(\ref{solution}) and~(\ref{solution.par}) have been obtained by considering $a=0$. A particular case with $a\ne 0$ can be obtained restricting Eq.~(\ref{solution}) for $x\ge 0$ and choosing, for instance, $r=s=1$, $a=a'-1$, $c=b'-1$ and $A=1$, where $a'>0$, $b'>0$. Thus,
\begin{equation}
  \rho(x,t)=\frac{x^{a'-1}}{B(a',b')t^{a'}}\dpar{1-\frac{x}{t}}^{b'-1}\,,\quad x\ge 0\,,
\end{equation}
which is a beta distribution, where $B(\cdot,\cdot)$ stands for the beta function. This function has been employed to model parliamentary presence and is a solution of the following diffusion equation~\cite{PhysRevE.99.042141}:
\begin{equation}
  \label{betaeq}
  \frac{\partial\rho}{\partial t}=D\frac{\partial}{\partial x}\dsqr{x^{a'}t^{a'/(b'-1)}\frac{\partial}{\partial x}\dpar{\frac{\rho}{x^{a'-1}}}^{b'/(b'-1)}}\,,
\end{equation}
where $D=(b')^{-1}[B(a',b')]^{1/(b'-1)}$.

In a more general framework, a diffusion equation that has the function $\rho(x,t)$, defined in Eq.~(\ref{solution}), as a solution is given by (see appendix~\ref{deduction}):
\begin{equation}
  \label{equation}
  \frac{\partial\rho}{\partial t}=D \frac{\partial }{\partial x}\dsqr{|x|^{a+\theta}t^\delta\frac{\partial }{\partial x}\dpar{\frac{\rho}{|x|^a}}^\nu}\,,
\end{equation}
where $\theta=2-r$, $\nu=1+1/c$, $\delta=s-1+(a+1)s/rc$ and
\begin{equation}
  \label{Dcoef}
  D=\frac{s}{AN^{1/c}(c+1)r^2}\,.
\end{equation}
We note immediately that the diffusion equations~(\ref{usual}), (\ref{Richardson}), (\ref{porous}) and~(\ref{betaeq}) are particular cases of Eq.~(\ref{equation}).

A more general scenario, which encompasses Eq.~(\ref{equation}), occurs if we consider the family of diffusion equations\begin{equation}
  \label{eq:general}
  \frac{\partial \rho}{\partial t}=\frac{\partial }{\partial x}\dcur{\mc D(x,t)\frac{\partial}{\partial x}\dsqr{\frac{\rho}{\mc C(x,t)}}^\nu}\,,
\end{equation}
where $\mc D(x,t)$ and $\mc C(x,t)$ are non-negative functions. In order to investigate stationary solutions and $H$ theorems related to this family of equations, we will focus our attention on the equation
\begin{equation}
\label{gfpe}
\frac{\partial\rho}{\partial t}=-\frac{\partial}{\partial x}[\mc A(x)\rho]+\frac{\partial}{\partial x}\dcur{\mc D(x) \frac{\partial}{\partial x}\dsqr{\frac{\rho}{\mc C(x)}}^\nu}\,,
\end{equation}
where $\mc A(x)$ is a drift coefficient and $\mc D(x)$ and $\mc C(x)$ are non-negative functions. In addition, we will consider Eq.~(\ref{gfpe}) in connection with probability distributions. Thus, Eq.~(\ref{gfpe}) will be interpreted as a generalized Fokker-Planck equation.

\section{Stationary solutions}
\label{sec:stat}

A stationary solution $\rho_s(x)$ of Eq.~(\ref{gfpe}) must satisfy the equation
\begin{equation}
\frac{d}{dx}\dcur{-\mc A(x)\rho_s+\mc D(x)\frac{d}{dx}\dsqr{\frac{\rho_s}{\mc C(x)}}^\nu} =0\,.
\end{equation}
Taking into account the reasonable assumption that $\rho_s(x)$ and its derivative converge to $0$ as $|x|\to+\infty$, we have that $\rho_s(x)$ must be a solution of the equation
\begin{equation}
  \label{eq:FP.stat0}
\mc D(x)\frac{d}{dx}\dsqr{\frac{\rho_s}{\mc C(x)}}^\nu=\mc A(x)\rho_s\,.
\end{equation}
If $\nu\ne 1$, this equation can be written as
\begin{equation}
  \label{eq:FP.stat1}
\frac{d}{dx}\dsqr{\frac{\rho_s}{\mc C(x)}}^{\nu-1}=\frac{\nu-1}{\nu}\frac{\mc C(x)\mc A(x)}{\mc D(x)}\,.
\end{equation}
Hence,
\begin{equation}
  \label{FP.stat}
\rho_s(x)=\mc C(x)\dsqr{K-\frac{\nu-1}{\nu}V(x)}_+^{1/(\nu-1)}\,,
\end{equation}
where $[x]_+=\max\{x,0\}$, $K$ is an integration constant that may be determined using a normalization condition and $V(x)$ is a function such that
\begin{equation}
  \label{V}
  \frac{dV}{dx}=-\frac{\mc C(x)\mc A(x)}{\mc D(x)}\,,
\end{equation}
which can be interpreted as an effective potential. Equation~(\ref{FP.stat}) can be rewritten as
\begin{equation}
  \label{eq:FP.stat.expq}
  \rho_s(x)=N\mc C(x)\exp_{2-\nu}\dsqr{-\frac{V(x)}{\nu N^{\nu-1}}}\,,
\end{equation}
where $N$ is a normalization constant and
\begin{equation}
  \exp_q(x)=e_q^x=
  \begin{cases}
    e^x&\text{for }q=1\\
    [1+(1-q)x]_+^{1/(1-q)}&\text{for }q\ne 1\,,
  \end{cases}
\end{equation}
which is usually referred as the $q$-exponential function.

If $\nu=1$, Eq.~(\ref{eq:FP.stat0}) can be reduced to the equation
\begin{equation}
  \frac{d}{dx}\ln\dsqr{\frac{\rho_s}{\mc C(x)}}=\frac{\mc C(x)\mc A(x)}{\mc D(x)}
\end{equation}
and, consequently,
\begin{equation}
  \label{eq:FP.stat.exp}
  \rho_s(x)=N\mc C(x)e^{-V(x)}\,,
\end{equation}
where $N$ is a normalization constant. Moreover, we note that Eq.~(\ref{eq:FP.stat.exp}) follows immediately from Eq.~(\ref{eq:FP.stat.expq}) if we take the limit $\nu\to 1$. Furthermore, Eqs.~(\ref{eq:FP.stat.expq}) and~(\ref{eq:FP.stat.exp}) give rise to the possibility of interpreting the function $\mc C(x)$ as a density of states.

In order to put in evidence the role of the effective potential $V(x)$, let us suppose that $\mc A(x)$ comes from a potential function $\phi(x)$, i.e. $\mc A(x)=-\phi'(x)$. In this case, the effective potential $V(x)$ would not be proportional to the original one $\phi(x)$ (except for an additive constant) unless the ratio $\mc C(x)/\mc D(x)$ is a constant. As an example, we can briefly discuss the particular case of Eq.~(\ref{gfpe}) with $\mc A(x)=-k|x|^{\lambda-1}x$, $k>0$, $\lambda\in\R$. If $\lambda=1$ and both $\mc C(x)$ and $\mc D(x)$ are constant functions, we can immediately verify that Eq.~(\ref{V}) implies that $V(x)$ is a harmonic potential. However, Eq.~(\ref{V}) shows that $V(x)$ can be very different from the harmonic potential if $\mc C(x)$ and $\mc D(x)$ are non-constant functions. On the other hand, if $\lambda\ne 1$, there are infinite possibilities for choosing the functions $\mc C(x)$ and $\mc D(x)$ such that the effective potential $V(x)$ is a harmonic one. For instance, if we consider $\mc D(x)=|x|^\alpha$ and $\mc C(x)=|x|^\beta$ with $\beta-\alpha=1-\lambda$, then we can obtain from Eq.~(\ref{V}) that $V(x)=kx^2/2$, where we have omitted a possible additive constant.

As we have remarked in section~\ref{sec.Eq}, Eq.~(\ref{gfpe}) can be seen as a large family of generalized Fokker-Planck equations that contains a broad spectrum of solutions. In particular, its stationary solution, given by Eq.~(\ref{eq:FP.stat.expq}), encompasses several well-known distributions, specially, all the probability distributions mentioned in the introduction. For instance, a $q$-Gaussian distribution can be obtained from~Eq.~(\ref{eq:FP.stat.expq}) by considering $\mc C(x)=1$ and $V(x)\propto x^2$. Moreover, if we consider $x\ge 0$, $\mc C(x)=x^a$ and $V(x)\propto x^r$ in Eq.~(\ref{eq:FP.stat.expq}), we obtain a generalized $q$-gamma distribution
\begin{equation}
  \rho_s(x)=Nx^ae_{2-\nu}^{-\beta x^r}\,,
\end{equation}
where $\beta$ is a constant proportional to $\nu^{-1}N^{1-\nu}$. This distribution includes Weibull ($a=r-1$ and $\nu\to 1$), gamma ($r=1$ and $\nu\to 1$) and $q$-gamma ($r=1$) distributions as particular cases. 

We stress that the stationary solution of Eq.~(\ref{gfpe}) is expressed in terms of the functions $\mc C(x)$ and $V(x)$, and the effective potential $V(x)$ depends on the ratio of $\mc C(x)\mc A(x)$ and $\mc D(x)$ (see Eq.~(\ref{V})). Hence, if the function $\mc C(x)$ is fixed, $V(x)$ can in principle be any function since we can conveniently choose the ratio $\mc A(x)/\mc D(x)$ in order to obtain the desired expression for the effective potential $V(x)$. As a consequence, different members of the family of equations~(\ref{gfpe}) can have the same stationary solution. Curiously, we note that Eq.~(\ref{gfpe}) may have a stationary solution even if we consider a null drift term. In fact, $\mc A(x)=0$ leads to a constant effective potential $V(x)$ and Eq.~(\Ref{eq:FP.stat.expq}) reduces to $\rho_s(x)=N\mc C(x)$. In this case, $\mc C(x)$ must be a normalizable function, which can be interpreted as a probability distribution. We can understand the peculiarity related to $\mc A(x)=0$ in Eq.~(\ref{gfpe}) noting that the solely presence of the function $\mc C(x)$ in this equation yields an unusual drift term. Indeed, Eq.~(\ref{gfpe}) can be written as
\begin{equation}
  \frac{\partial\rho}{\partial t}=-\frac{\partial}{\partial x}[\tilde{\mc A}(x,\rho)\rho]+\frac{\partial}{\partial x}\dsqr{\tilde{\mc D}(x)\frac{\partial\rho^\nu}{\partial x}}\,,
\end{equation}
where
\begin{equation}
  \label{eq:Atilde}
  \tilde{\mc A}(x,\rho)=\mc A(x)+\nu\frac{\tilde{\mc D}(x)}{\mc C(x)}\frac{d \mc C(x)}{dx}\,\rho^{\nu-1}
\end{equation}
and
\begin{equation}
  \tilde{\mc D}(x)=\frac{\mc D(x)}{[\mc C(x)]^\nu}\,.
\end{equation}
Thus, we identify the second term on the right hand side of Eq.~(\ref{eq:Atilde}) as the referred  unusual drift term, which depends on $\rho(x,t)$ if $\nu\ne 1$. On the other hand, $\tilde{\mc D}(x)$ can be viewed as an effective diffusion coefficient.



\section{$H$ theorem}
\label{sec:Hthm}

As it has been done for other generalized Fokker-Planck equations~\cite{SchwammleCuradoNobre2007,SchwammleNobreCurado2007,PhysRevE.94.062117}, an $H$ theorem can be verified considering a free-energy-like functional $F=U-S$. In our case, we can verify this theorem for Eq.~(\ref{gfpe}) using
\begin{equation}
\label{U}
  U=\int_{-\infty}^{+\infty} V(x)\rho\,dx
\end{equation}
and
\begin{equation}
  \label{S}
  S=\int_{-\infty}^{+\infty} \frac{[\mc C(x)]^{1-\nu}\rho^\nu-\rho}{1-\nu}\,dx\,.
\end{equation}
The functional $U$ would play the role of an internal energy unless by a multiplicative constant, possibly related to a generalized temperature. On the other hand, the functional $S$ can be viewed as an entropic-like form.

The time derivative of $F$ is given by
\begin{equation}
  \frac{dF}{dt}=\int_{-\infty}^{+\infty} \dcur{V(x)-\frac{[\mc C(x)]^{1-\nu}\nu\rho^{\nu-1}-1}{1-\nu}}\frac{\partial \rho}{\partial t}\,dx\,.
\end{equation}
Using Eq.~(\ref{gfpe}), we have
\begin{equation}
  \begin{split}
    \frac{dF}{dt}&=\int_{-\infty}^{+\infty} \dcur{V(x)-\frac{[\mc C(x)]^{1-\nu}\nu\rho^{\nu-1}-1}{1-\nu}}\\
    &\quad\times\frac{\partial}{\partial x}\dcur{-\mc A(x)\rho+\mc D(x)\frac{\partial }{\partial x}\dsqr{\frac{\rho}{\mc C(x)}}^\nu}\,dx\,.
  \end{split}
\end{equation}
Then, integrating by parts and assuming that $\rho(x,t)$ approaches $0$ rapidly enough as $|x|$ increases beyond all bounds, we have
\begin{equation}
  \begin{split}
    \frac{dF}{dt}&=-\int_{-\infty}^{+\infty} \dcur{-\frac{\mc C(x)\mc A(x)}{\mc D(x)}+\frac{\nu}{\nu-1}\frac{\partial}{\partial x}\dsqr{\frac{\rho}{\mc C(x)}}^{\nu-1}}\\
    &\quad\times\dcur{-\mc A(x)\rho+\mc D(x)\frac{\partial }{\partial x}\dsqr{\frac{\rho}{\mc C(x)}}^\nu}\,dx\\
    &=-\int_{-\infty}^{+\infty}\frac{\mc D(x)\rho}{\mc C(x)}\left\{-\frac{\mc C(x)\mc A(x)}{\mc D(x)}\vphantom{\dsqr{\frac{\rho}{\mc C(x)}}^{\nu-1}}\right.\\
    &\quad\left.+\frac{\nu}{\nu-1}\frac{\partial}{\partial x}\dsqr{\frac{\rho}{\mc C(x)}}^{\nu-1}\right\}^2\,dx\,.
  \end{split}
\end{equation}
Therefore, since $\rho(x,t)$, $\mc D(x)$ and $\mc C(x)$ are assumed to be non-negative functions, we have
\begin{equation}
  \label{Hthm}
  \frac{dF}{dt}\le 0\,,
\end{equation}
which can be seen as an $H$ theorem associated with Eq.~(\ref{gfpe}).

In addition to the last result, by virtue of Eq.~(\ref{FP.stat}), the effective potential $V(x)$ can be written in terms of the stationary solution $\rho_s(x)$ as
\begin{equation}
V(x)=\frac{\nu}{\nu-1}\dcur{K-\dsqr{\frac{\rho_s}{\mc C(x)}}^{\nu-1}}\,.
\end{equation}
Using this in Eq.~(\ref{U}), we obtain that the free-energy-like functional $F$ is given by
\begin{equation}
\begin{split}
F(\rho)&=\frac{\nu K-1}{\nu-1}\\
&\quad+\int_{-\infty}^{+\infty}\frac{1}{[\mc C(x)]^{\nu-1}}\dpar{\frac{\rho^\nu}{\nu-1}-\frac{\nu\rho_s^{\nu-1}\rho}{\nu-1}}\,dx\,,
\end{split}
\end{equation}
where we have used the normalization condition $\int_{-\infty}^{+\infty}\rho\,dx=1$. Hence,
\begin{equation}
  \label{DeltaF1}
  F(\rho)-F(\rho_s)=\int_{-\infty}^{+\infty} \frac{\rho_s^\nu}{[\mc C(x)]^{\nu-1}}\,g\dpar{\frac{\rho}{\rho_s}}\,dx\,,
\end{equation}
where
\begin{equation}
  g(z)=\frac{z^\nu-\nu z}{\nu-1}+1\,,\quad z\ge 0\,.
\end{equation}
Considering $\nu>0$, we note that the function $g(z)$ has a global minimum at the point $1$ and, consequently, $g(z)\ge g(1) =0$ for all $z\ge 0$. Using this fact in Eq.~(\ref{DeltaF1}), we obtain
\begin{equation}
  \label{DeltaF}
  F(\rho)\ge F(\rho_s)\,.
\end{equation}

The essence of the $H$ theorem associated with Eq.~(\ref{gfpe}) is that, if we assume that $\rho_s(x)$ is the only stationary solution of Eq.~(\ref{gfpe}) for $\nu>0$, by virtue of the inequalities~(\ref{Hthm}) and~(\ref{DeltaF}), every solution of Eq.~(\ref{gfpe}) tends to $\rho_s(x)$ as $t$ increases without bound, i.e. $\rho_s(x)$ is the equilibrium solution. 


The entropic-like form $S$ given in Eq.~(\ref{S}), which has an unusual dependence on the position, can be seen as a generalization of the Tsallis entropic form~\cite{Tsallis1988}, 
\begin{equation}
  \label{Sq}
  S_\nu=\int_{-\infty}^{+\infty}\frac{\rho^{\nu}-\rho}{1-\nu}
\,dx\,,
\end{equation}
since Eq.~(\ref{Sq}) can be obtained from Eq.~(\ref{S}) by taking $\mc C(x)=1$. In particular, if $\mc C(x)=1$ and $\nu\to 1$ we recover the Shannon entropic form. These entropic forms can be compactly written employing the $q$-logarithmic function 
\begin{equation}
  \ln_qx=\begin{cases}
    \ln x&\text{for }q=1\\
    \frac{x^{1-q}-1}{1-q}&\text{for }q\ne 1\,.
  \end{cases}
\end{equation}
Indeed, using this function, we have
\begin{equation}
  \label{S.1}
  S=-\int_{-\infty}^{+\infty} \rho\ln_{2-\nu}\dsqr{\frac{\rho}{\mc C(x)}}\,dx\,.
\end{equation}

In statistical mechanics, a well known procedure to obtain equilibrium distributions consist in optimizing the entropy taking into account some constraints. This procedure can also be implemented considering other entropic forms; for instance, in the study of complex systems. To illustrate this approach, we consider here the optimization of the entropic form given in Eq.~(\ref{S.1}) subjected to the constraints
\begin{equation}
  \int_{-\infty}^{+\infty} \rho\,dx=1\quad\text{and}\quad\int_{-\infty}^{+\infty} \rho \tilde V(x)\,dx=U\,.
\end{equation}
In this direction, we consider the functional
\begin{equation}
  \begin{split}
    L&=\int_{-\infty}^{+\infty}\rho\ln_{2-\nu}\dsqr{\frac{\rho}{\mc C(x)}}\,dx-\tilde\alpha\dpar{\int_{-\infty}^{+\infty}\rho\,dx-1}\\
    &\quad-\tilde\beta\dpar{\int_{-\infty}^{+\infty}\rho \tilde V(x)\,dx-U}\,,
\end{split}
\end{equation}
where $\tilde\alpha$ and $\tilde \beta$ are Lagrange multipliers. The solution of the equation $\delta L/\delta \rho=0$ leads to the generalized Boltzmann-Gibbs distribution
\begin{equation}
  \rho(x)=N\mc C(x)\exp_{2-\nu}[-\beta\tilde V(x)]\,,
\end{equation}
where $N$ is a normalization constant and $\beta$ is a parameter related to the Lagrange multipliers. We note that this distribution coincides with the stationary solution given in Eq.~(\ref{eq:FP.stat.expq}) if $\tilde V(x)$ is equal to the effective potential $V(x)$. Moreover, if we only consider the normalization constraint, we obtain $\rho(x)=N\mc C(x)$, provided that $\mc C(x)$ can be normalized. This result recovers the case of equiprobability when $\mc C(x)$ is a constant function on a compact interval.

As a final remark about the representation of the entropic form $S$ given in Eq.~(\ref{S.1}), we note that $-S$ can be formally seen as a generalized relative entropy~\cite{Tsallis1998}. In fact, if $\nu\to 1$, $-S$ reduces to the usual form of the relative entropy, which is also known as the Kullback-Leibler divergence. In this context, $\mc C(x)$ should be thought as a probability distribution, which could be also seen as a normalized density of states in connection with Eq.~(\ref{eq:FP.stat.expq}).

The terms in the free-energy-like functional $F$ can be arranged in a different manner considering other definitions for its internal energy and entropic form parts. In this direction, we can write $F=\mc U-S_\nu$, where $S_\nu$ is the Tsallis entropic form given in Eq.~(\ref{Sq}) and
  \begin{equation}
    \mc U=\int_{-\infty}^{+\infty}[\rho V(x)-\rho^\nu\ln_\nu \mc C(x)]\,dx\,.
  \end{equation}
  We note that if $\mc C(x)=1$ for every $x$, the two forms of writing the functional $F$, namely as $U-S$ or $\mc U-S_\nu$, become identical. For other definitions of the function $\mc C(x)$, the functionals $\mc U$ and $U$ differ by the term $-\int_{-\infty}^{+\infty}\rho^\nu \ln_\nu \mc C(x)\,dx$, which seems like a mean of the ``potential'' $-\ln_\nu \mc C(x)$ weighted by $\rho^\nu$. This kind of mean has been employed in several works related to the nonextensive statistical mechanics~\cite{Tsallis1988,Tsallis}.

The two ways of decomposing the functional $F$ discussed here indicates the necessity of the introduction of unusual terms. If we retain a conventional internal energy part, the remaining terms in $F$ compose an unusual entropic form, depending on the position. On the other hand, if we consider a Tsallis entropic form in the functional $F$, we are led to an unconventional internal energy, involving a pseudo-mean which does not return $1$ when it is applied to $1$. Nevertheless, the optimization of the Tsallis entropic form subjected to the constraints $\int_{-\infty}^{+\infty}\rho\,dx=1$ and $\mc U=\text{constant}$ leads to a distribution which is different from the stationary solution given in Eq.~(\ref{eq:FP.stat.expq}). In fact, in this case we obtain
  \begin{equation}
    \rho(x)=Ne_\nu^{\tilde\beta \ln_\nu\mc C(x)}e_{2-\nu}^{-\beta V(x)}\,,
  \end{equation}
  where $\beta$ is a constant related to the Lagrange multiplier $\tilde\beta$ and $N$ is a normalization constant.
  This indicates that the manner of decomposing the functional $F$ as composed by an internal energy and an entropic form should be carefully thought. In particular, our options of decomposition are restricted if we desire to have an agreement between the stationary solution and the distribution that optimizes the entropic form. In this direction, the decomposition $F=U-S$ seems to be a better choice than $F=\mc U-S_\nu$.

\section{Conclusions}
\label{sec:conc}


We have investigated a broad family of generalized Fokker-Planck equations that contains Richardson~\cite{doi:10.1098/rspa.1926.0043} and porous media equations~\cite{PlastinoPlastino1995,PhysRevE.94.062117} as particular members (see Eq.~(\ref{gfpe})). We have found for each equation of this family a stationary solution considering that the drift term $\mc A(x)$ is confining. 
Moreover, this stationary solution only depends on the functions $\mc C(x)$ and an effective potential $V(x)$ (see Eq.~(\ref{eq:FP.stat.expq})). Since $V(x)$ depends on the the ratio of $\mc C(x)\mc A(x)$ and $\mc D(x)$, different members of the family of equations~(\ref{gfpe}) can have the same stationary solution. Curiously, if $V(x)$ is a constant function, Eq.~(\ref{gfpe}) may still have a stationary solution, which is proportional to $\mc C(x)$. Another characteristic of the stationary solutions is that a large set of well-known distributions can be obtained as particular cases; for instance, Weibull, gamma and $q$-gamma distributions.

In addition to finding stationary solutions for the proposed family of generalized Fokker-Planck equations, given in Eq.~(\ref{gfpe}), we have verified an $H$ theorem for each member of this family. The $H$ theorem allows us to say that, if a generalized Fokker-Planck equation has a unique stationary solution $\rho_s(x)$, then any (well-behaved) solution of this equation tends to $\rho_s(x)$ as time increases without bound. Thus, $\rho_s(x)$ turns out to be the equilibrium solution. In order to verify the $H$ theorem, we have considered a free-energy-like functional $F=U-S$, as done in several articles on this subject~\cite{SchwammleCuradoNobre2007,SchwammleNobreCurado2007,PhysRevE.94.062117}. In our study, $U$ is the average of the effective potential $V(x)$ and $S$ is an entropic-like form that generalizes the Tsallis one~\cite{Tsallis1988,Tsallis}. The optimization of this entropic-like form taking into account a normalization condition and a constant value for $U$ also yields the stationary solution $\rho_s(x)$. A further fact about this entropic-like form $S$ (see Eq.~(\ref{S})) is that it can depend explicitly on the position since it contains the function $\mc C(x)$ in its definition. Remarkably, $-S$ has the form of a generalized Kullback-Leibler divergence, indicating the possibility of further investigations or applications out of the context of $H$ theorem and Fokker-Planck equations; for instance, in the study of complex systems.

We have also shown that the free-energy-like functional $F$ is consistent with different definitions of internal energy and entropic form.
   In particular, we have exhibited two ways of decomposing the functional $F$, which reveal the necessity of considering unusual terms. In fact, we need to consider an unusual entropic form, depending on the position, or an unconventional internal energy, involving a pseudo-mean which does not return $1$ when it is applied to $1$. However, the use of the latter in the optimization of the Tsallis entropic form leads to a distribution which is not consistent with the stationary solution $\rho_s(x)$. Thus, the manner of decomposing
the functional $F$ as composed by an internal energy and an
entropic form can induce to undesirable results.

More general situations than the one discussed in this work can be conducted considering, for instance, that the functions $\mc D$ and $\mc C$ depend explicitly on the distribution $\rho$ in addition to the position. In this case we can expect that other effective potentials and other relative entropic forms may emerge. A more ambitious study of these two aspects could be based on generalizations of the Fokker-Planck equation additionally involving fractional derivatives. Furthermore, in principle, all these possible investigations could be also extended for more than one dimension.

\acknowledgements

The authors thank CNPq and CAPES (Brazilian funding agencies) for partial financial support. 

\appendix

\section{An elementary deduction of Eq.~(\ref{equation})}
\label{deduction}
We will briefly describe how to obtain a diffusion equation that has the distribution $\rho(x,t)$, defined in Eq.~(\ref{solution}), as a solution. As a starting point, we consider the equation
\begin{equation}
  \label{eq:Ervin}
\frac{\partial \tilde\rho}{\partial \tau}=\tilde D\frac{\partial}{\partial x}\dpar{|x|^\theta\frac{\partial \tilde\rho^\nu}{\partial x}}\qquad (\nu\ne 1)\,.
\end{equation}
A solution for this equation is the function~\cite{PhysRevE.63.030101}:
\begin{equation}
  \label{sol:Ervin}
  \tilde\rho(x,\tau)=\frac{N}{\tau^{c/(rc+1)}}\dpar{1-A\frac{|x|^r}{\tau^{rc/(rc+1)}}}_+^c\,,
\end{equation}
where $r=2-\theta$, $c=1/(\nu-1)$ and $A$ and $N$ are constants satisfying the relation
\begin{equation}
  \tilde D=\frac{rc}{AN^{1/c}(c+1)(rc+1)r^2}\,.
\end{equation}
If $\tau=t^{\gamma+1}$ and $\tilde\rho(x,\tau)=\rho_0(x,t)$, then Eq.~(\ref{eq:Ervin}) reads
\begin{equation}
  \label{eq:inter}
  \frac{\partial \rho_0}{\partial t}=D\frac{\partial }{\partial x}\dpar{|x|^\theta t^\gamma\frac{\partial \rho_0^\nu}{\partial x}}\,,
\end{equation}
where $D=(\gamma+1)\tilde D$. In addition, Eq.~(\ref{sol:Ervin}) assumes the form
\begin{equation}
  \rho_0(x,t)=\frac{N}{t^{s/r}}\dpar{1-A\frac{|x|^r}{t^s}}_+^c\,,
\end{equation}
where, $r=2-\theta$, $c=1/(\nu-1)$, $s=(\gamma+1)rc/(rc+1)$ and $A$ and $N$ are constants connected with $D$ by Eq.~(\ref{Dcoef}). The function $\rho(x,t)$, defined by Eq.~(\ref{solution}), is related to the function $\rho_0(x,t)$ by the equation
\begin{equation}
  \rho(x,t)=\frac{|x|^a}{t^{as/r}}\rho_0(x,t)\,.
\end{equation}
Then, using Eq.~(\ref{eq:inter}), the following differential equation is certainly satisfied:
\begin{equation}
  \frac{\partial}{\partial t}\dpar{\frac{t^{as/r}}{|x|^a}\rho}=D\frac{\partial }{\partial x}\dsqr{|x|^\theta t^\gamma\frac{\partial}{\partial x}\dpar{\frac{t^{as/r}}{|x|^a}\rho}^\nu}\,.
\end{equation}
Employing the product rule of differentiation on the time derivative and on the outer spatial derivative, we obtain
\begin{equation}
  \frac{as}{rt}\rho+\frac{\partial\rho}{\partial t}=Dt^\delta |x|^a\frac{\partial }{\partial x}\dsqr{|x|^\theta\frac{\partial }{\partial x}\dpar{\frac{\rho}{|x|^a}}^\nu}\,,
\end{equation}
where $\delta=\gamma+(\nu-1)as/r$. Then,
\begin{equation}
  \label{eq:A9}
  \begin{split}
    \frac{as}{rt}\rho+\frac{\partial\rho}{\partial t}&=-aDt^\delta |x|^{a+\theta-1}\frac{\partial }{\partial x}\dpar{\frac{\rho}{|x|^a}}^\nu\\
    &\quad+Dt^\delta \frac{\partial }{\partial x}\dsqr{|x|^{a+\theta}\frac{\partial }{\partial x}\dpar{\frac{\rho}{|x|^a}}^\nu}\,.
  \end{split}
\end{equation}
Using Eq.~(\ref{solution}), we can verify that the first terms on both sides of Eq.~(\ref{eq:A9}) are identical. Therefore, the function $\rho(x,t)$, defined by Eq.~(\ref{solution}), is a solution of Eq.~(\ref{equation}).

\bibliographystyle{apsrev4-2}
\bibliography{biblio}

\end{document}